\documentclass{IEEEtran}
\usepackage{graphicx}
\usepackage{amsmath, amssymb, mathtools, stmaryrd, bm, gensymb} % math packages
\usepackage{url}
\usepackage{tikz, pgfplots}
\usepackage{siunitx}
\usepackage{algorithm, algpseudocode}	% tabela de algoritmos
\usepackage{siunitx}
\usepackage{cite}
\usepackage{flushend}

\mathtoolsset{showonlyrefs} %mathtools

\usetikzlibrary{plotmarks}

\newcommand{\mbb}[1]{\ensuremath\mathbb{#1}}
\newcommand{\mc}[1]{\ensuremath{\mathcal{#1}}}				% calligraphic letters
\newcommand{\tran}{\mathsf{T}}						% transpose operator
\newcommand{\hermit}{\mathsf{H}}					% hermitian operator
\newcommand{\neuclid}[1]{\ensuremath{\left\|#1\right\|_\textrm{2}}}	% l2 norm
\newcommand{\esp}[1]{\ensuremath{\mathbb{E}\left[#1\right]}}		% stat expectation

\begin{document}
	
	\title{Low-Rank Tensor MMSE Equalization}
	
	\pagenumbering{gobble}
	
\author{\IEEEauthorblockN{
		% authors
		Lucas N. Ribeiro,
		Andr\'e L. F. de Almeida,
		Jo\~ao C. M. Mota
	}

\thanks{This work is partially supported by the Brazilian National Council for Scientific and Technological Development - CNPq, CAPES/PROBRAL Proc. numbers 88887.144009/2017-00, 308317/2018-1, and FUNCAP.}
	
	% marks 
	\IEEEauthorblockA{Wireless Telecommunications Research Group (GTEL), Universidade Federal do Cear\'a, Fortaleza, Brazil}

	% emails
	Emails: \{nogueira, andre, mota\}@gtel.ufc.br

}
	\maketitle
	
	\begin{abstract}
        New-generation wireless communication systems will employ large-scale antenna arrays to satisfy the increasing capacity demand. This massive scenario brings new challenges to the channel equalization problem due to the increased signal processing complexity. We present a novel low-rank tensor equalizer to tackle the high computational demands of the classical linear approach. Specifically, we propose a method to design a canonical polyadic tensor filter to minimize the mean square error criterion. Our simulation results indicate that the proposed equalizer needs fewer calculations and is more robust to short training sequences than the benchmark.
	\end{abstract}
	
	
	\section{Introduction}
	
	Modern wireless communication systems rely on large-scale antenna arrays to enhance their performance~\cite{larsson_massive_2014,schwarz_society_2016}. Such massive arrays yield high beamforming gain, improve interference suppression and ameliorate the spatial resolution capabilities of the system. However, the implementation of large-scale arrays raises some challenges, including computationally demanding signal processing, high energy consumption, among others. Tensor filtering has been investigated as a possible solution to the high computational complexity issue of large-scale systems~\cite{ribeiro_tensor_2016,ribeiro_iet18,ribeiro2019separable,paleologu18,elisei2018efficient,cohen2019differential,rupp_tensor_2015,rupp_gradient-based_2015,liu2018robust}.
	
    In~\cite{ribeiro_tensor_2016,ribeiro_iet18,ribeiro2019separable}, we present beamforming methods for massive arrays considering a modest Kronecker separable system model. Therein, the high-dimensional beamforming vector is factorized into a Kronecker product of lower-dimension factors. Such a factorization allows us to optimize the beamformer for each low-dimensional factor, decreasing the number of calculations. We also observe that Kronecker beamformers may drastically reduce the number of calculations in the beamforming optimization with negligible signal recovery performance degradation. These filters, however, do not provide many degrees of freedom due to their rank-$1$ structure, limiting the performance and the applicability to more practical system models.
    
    A general formulation of Kronecker filters is thus necessary to enhance their performance. In fact, Kronecker separable filters can be regarded as rank-$1$ tensor filters~\cite{comon2014tensors,kolda_tensor_2009}. One way to increase the filter's degrees of freedom consists of employing low-rank filters, i.e., we consider a finite sum of Kronecker-separable terms. In~\cite{paleologu18,elisei2018efficient}, low-rank bilinear system identification methods are proposed to estimate acoustic impulse responses. It is shown that some sparse acoustic signatures nicely fits the low-rank bilinear model. In fact, sparsity is strongly linked to the low-rank system representation~\cite{sparse}. Some works~\cite{ribeiro_identification_2015,dogariu2019system} also consider the identification of trilinear systems. These estimation methods exhibit better accuracy than their classical counterparts. In the context of wireless communications, \cite{filiz_rank_2004} presents a low-rank bilinear filter for code division multiple access systems. The influence of the filter rank on the system performance is studied. It is shown that the rank parameter controls a complexity-equalization performance trade-off. Unfortunately, the analysis of~\cite{filiz_rank_2004} is restricted to the bilinear case and details on its computational complexity are lacking. Therefore the potential of multilinear filters is yet to be investigated.
	
	The main contributions of the present paper can be summarized as:
	(i) We propose a novel low-rank multilinear equalizer for large-scale antenna array system based on the minimum mean square error (MMSE) filter. Our method extends those of \cite{ribeiro_tensor_2016,ribeiro_iet18,ribeiro2019separable} to deal with non-separable systems and it also generalizes those of \cite{paleologu18,dogariu2019system,filiz_rank_2004} to the multidimensional case; (ii) We investigate the computational complexity and the equalization performance of the proposed method; (iii) With simulation results, we demonstrate the robustness of our method to short training sequences and its superior computational efficiency compared to the classical linear equalization approach.
	
	\subsection{Notation}
	$\bm{x}$ denotes vectors, $\bm{X}$ matrices and $\mc{X}$ tensors. $[\bm{X}]_{i,j}$ stands for the $(i,j)$-th entry of $\bm{X}$. The transpose, and the conjugate transpose (Hermitian) of $\bm{X}$ are denoted by $\bm{X}^\tran$ and $\bm{X}^\hermit$, respectively. The $(M\times M)$-dimensional identity matrix is represented by $\bm{I}_M$ and the $(M\times N)$-dimensional null matrix by $\bm{0}_{M\times N}$. The $\ell_2$ norm, the statistical expected value operator and the vectorization operator are respectively denoted as $\neuclid{\cdot}$, $\esp{\cdot}$, $\text{vec}(\cdot)$. The outer product, the Kronecker product, the $n$-mode product and the Big-O notation are referred to as $\circ$, $\otimes$, $\times_n$ and $O(\cdot)$, respectively.
	
	\section{System Model}
	Let us consider a multiple-input multiple-output (MIMO) wireless communication system consisting of $U$ user equipment (UE) and a single base-station (BS). We assume that each UE is equipped with a single omni-directional antenna and the BS employs a uniform linear array (ULA) of $N$ omni-directional antennas whose axis is parallel to the ground plane. The spacing between the array  antennas is considered to be $d=\lambda/2$, where $\lambda$ denotes the carrier wavelength. This half-wavelength ULA setup is considered for simplicity purposes, however, our model can be easily adapted to different kinds of antenna geometry. We consider the uplink scenario, where UE $u$ emits an independent and identically distributed (i.i.d.) discrete-time digitally-modulated symbol sequence  $s_u[k]$ with zero mean and variance $\sigma_s^2$, where $k$ denotes the symbol period for $u \in \{1,\ldots,U\}$. From our assumptions, it follows that
	\begin{equation}
	\esp{s_i[k-p]s_j^*[k-q]} = \begin{cases}
	0, & i\neq j\text{ or } p\neq q\\
	\sigma_s^2, & i=j \text{ and } p = q
	\end{cases}.
	\end{equation}
	We assume a frequency-selective wireless channel with $Q$ delay taps and $L$ multi-paths. Therefore, the discrete-time representation of the received signal at the $n$-th BS antenna can be expressed as~\cite{Almers2007}
	\begin{equation}
	x_n[k] = \sum_{u=1}^U \sum_{q=0}^{Q-1} \sum_{\ell=0}^{L-1} \alpha_{u,\ell} g(qT - \tau_{u,\ell}) a_n(\theta_{u,\ell}) s_u[k-q] + b_n[k],
	\end{equation}
	where $\alpha_{u,\ell}$ denotes the complex channel gain, $g(\cdot)$ the effective pulse-shaping waveform, $\tau_{u,\ell}$ the propagation delay,  $T$ the symbol period, $a_n(\theta_{u,\ell})$ the channel spatial response and $b_n[k]$ the additive white Gaussian noise (AWGN) component for $n \in \{1,\ldots,N\}$. The channel gains are modeled as i.i.d. Gaussian random variables with zero mean and unit variance, and the AWGN components are Gaussian random variables with zero mean and $\sigma_n^2$ variance. Since the BS employs a half-wavelength ULA, the channel spatial response term is given by $a_n(\theta_{u,\ell}) = e^{-\jmath\pi(n-1)\cos(\theta_{u,\ell})}$, where $\theta_{u,\ell}$ stands for the direction of arrival of the $\ell$-th path associated with UE $u$. We also define the signal to noise ratio (SNR) as $\text{SNR} = \sigma_s^2/\sigma_n^2$.
	
	Let $\bm{x}[k] = [x_1[k], \ldots, x_N[k]]^\tran$ denote the received signal vector at BS. It can be written as:
	\begin{gather}
	\bm{x}[k] = \sum_{u=1}^U\bm{H}_u \bm{s}_u[k] + \bm{b}[k],\label{eq:rx}\\
	\bm{s}_u[k] = \left[ s_u[k], \ldots, s_u[k-Q+1] \right]^\tran, \\
	\bm{b}[k] = \left[ b_1[k], \ldots, b_N[k] \right]^\tran,
	\end{gather}
	where
	\begin{gather}
	\bm{H}_u = \sum_{\ell = 1}^L \alpha_{u,\ell} \bm{a}(\theta_{u,\ell}) \bm{g}(\tau_{u,\ell})^\tran \in \mbb{C}^{N\times Q}, \label{eq:chan}\\
	\bm{a}(\theta_{u,\ell}) = \left[1, \ldots, e^{-\jmath\pi(N-1)\cos(\theta_{u,\ell})}\right]^\tran \in \mbb{C}^{N},\\
	\bm{g}(\tau_{u,\ell}) = \left[ g(-\tau_{u,\ell}), \ldots, g((Q-1)T-\tau_{u,\ell})\right]^\tran \in \mbb{C}^{Q}
	\end{gather}
	denote the BS-UE $u$ uplink channel matrix, the array steering vector (spatial response), and the effective pulse-shaping vector (temporal response), respectively. Model~\eqref{eq:rx} assumes that the channel is block-fading, i.e., $\bm{H}_u$ remains constant over a frame of $K$ symbol periods.
	
	Let us consider that the BS wishes to extract from $\bm{x}[k]$ the symbols of UE $u$ while regarding the signals from the other $(U-1)$ users as interference. To emphasize this scenario, we rewrite \eqref{eq:rx} as
	\begin{equation} \label{eq:rx_int}
	\bm{x}[k] = \bm{H}_u \bm{s}_{u}[k] + \sum_{j\neq u}^U \bm{H}_{j} \bm{s}_j[k] + \bm{b}[k] 
	\end{equation}
	where the first term corresponds to the desired signal and the other terms to  interference and noise. The covariance matrix of \eqref{eq:rx_int} is defined as $\bm{R}_{xx} =\esp{\bm{x}[k]\bm{x}^\hermit[k]}$. From our assumptions, the covariance matrix can be written as $\bm{R}_{xx} = \bm{R}_{dd} + \bm{R}_{ii} + \bm{R}_{bb}$, where $\bm{R}_{dd} = \bm{H}_u \bm{R}_{ss} \bm{H}_u^\hermit$, $\bm{R}_{ii} = \sum_{j\neq u}^{U} \bm{H}_j \bm{R}_{ss} \bm{H}_j^\hermit$, $\bm{R}_{ss} = \sigma_s^2 \bm{I}_Q$, $\bm{R}_{bb} = \sigma_n^2 \bm{I}_{N}$.
	
	\section{Equalization Methods}

	We consider an equalizer filter to extract the desired signal from the observed signals and to compensate for channel distortions, multi-user interference, and noise. Let $\bm{w} \in \mbb{C}^{N}$ represent the equalizer weights vector. The filter output can thus be written as $y[k] = \bm{w}^\hermit \bm{x}[k]$. We are interested in designing the equalizer weights to minimize the difference between filter output and the desired symbol sequence (lagged by some $\delta$ due to the propagation delay). We consider the mean square error (MSE) as an optimization metric, therefore the equalizer design problem can be stated as
	\begin{equation} \label{eq:mse}
	\min_{\bm{w}}\, \esp{|s_u[k-\delta] - y[k]|^2}.
	\end{equation}
	
	In the following, we recall the classical MMSE filter, which is the optimal solution to \eqref{eq:mse}. We discuss some of its properties and drawbacks. Next, we propose a novel equalizer design method based on tensor algebra that addresses some issues of the classical MMSE equalizer.
	
	\subsection{Linear MMSE Filter} \label{sec:mmse}
	
	Let $J(\bm{w}) = \esp{|s_u[k-\delta] - y[k]|^2}$ denote the MSE objective function. It can be rewritten as
	\begin{equation}
	J(\bm{w}) = \sigma_s^2 - \bm{p}^\hermit\bm{w} - \bm{w}^\hermit\bm{p} + \bm{w}^\hermit \bm{R}_{xx}\bm{w}, \label{eq:msef}
	\end{equation}
	where $\bm{p} = \esp{\bm{x}[k]s_u^*[k-\delta]} = \bm{H}_u \bm{R}_{ss} \bm{e}_\delta$ is the cross-covariance vector between the elements of $\bm{x}[k]$ and $s_u[k-\delta]$, and $\bm{e}_\delta = [0, \ldots, 1, \ldots, 0]^\tran$ is a $Q$-dimensional vector with $1$ at its $\delta$-th entry and $0$ elsewhere. Since $J(\bm{w})$ is convex, its global minimizer can be found by solving $\nabla J(\bm{w}) = \frac{\partial J}{\partial \bm{w}^*} = \bm{0}_{N\times1}$. The MMSE filter is then given by
	\begin{align} \label{eq:mmsefilt}
	\bm{w}_{\text{MMSE}} &= \bm{R}_{xx}^{-1} \bm{p}.
	\end{align}
	
	The minimum MSE is obtained by substituting \eqref{eq:mmsefilt} into \eqref{eq:msef}:
	\begin{align}
	J(\bm{w}_\text{MMSE}) &= \sigma_s^2 - \bm{p}^\hermit\bm{R}_{xx}^{-1}\bm{p}\\
	&= \sigma_s^2 - \bm{e}_\delta^\tran \bm{R}_{ss} \bm{H}_u^\hermit \bm{R}_{xx}^{-1} \bm{H}_u \bm{R}_{ss} \bm{e}_\delta \label{eq:min_mse}.
	\end{align}
	Equation~\eqref{eq:min_mse} reveals that the choice of $\delta$ determines the minimum of the objective function. Therefore, to minimize~\eqref{eq:min_mse} with respect to $\delta$, one has to simply select $\delta$ as the index of the largest diagonal element of $\bm{R}_{ss} \bm{H}_u^\hermit \bm{R}_{xx}^{-1} \bm{H}_u \bm{R}_{ss} $.
	
	The \emph{a priori} knowledge of $\bm{R}_{xx}$ and $\bm{p}$ is hardly practical, and, thus, the statistics need to be estimated. In this case, one can consider a $K$-length training symbol sequence. Define  $\bm{X} = \left[ \bm{x}[0],\ldots,\bm{x}[K-1] \right] \in \mbb{C}^{N \times K}$ and $\bm{s}_u = \left[ s_u[-\delta],\ldots, s_u[K-1-\delta] \right]^\tran \in \mbb{C}^{K}$. The sample estimates are then given by 
	\begin{gather}
	\bm{R}_{xx} \approx \frac{1}{K} \sum_{k=0}^{K-1} \bm{x}[k]\bm{x}^\hermit[k] = \frac{1}{K} \bm{X}\bm{X}^\hermit, \label{eq:estRxx}\\
	\bm{p} \approx \frac{1}{K} \sum_{k=0}^{K-1} \bm{x}[k] s_u^*[k-\delta] = \frac{1}{K} \bm{X} \bm{s}_u^*. \label{eq:estP}
	\end{gather}
	
	Unfortunately, the MMSE filter faces some issues when sample estimates are considered in the large-scale scenario. Long training symbol sequences are necessary to obtain sufficiently accurate statistics due to the large dimension of $\bm{R}_{xx}$. Moreover, the number of computations in \eqref{eq:mmsefilt} may be exceedingly large. To be more specific, one needs to carry out $N^2K + NK$ products to estimate \eqref{eq:estRxx} and \eqref{eq:estP}, and $O(N^3) + N^2$ products to calculate \eqref{eq:mmsefilt}, which yields a total of
	\begin{equation}
	P_\text{MMSE}(N, K) = N^2K + NK + O(N^3) + N^2 \label{eq:count_mmse}
	\end{equation}
	products. For large arrays, the cubic and quadratic terms in~\eqref{eq:count_mmse} yields a substantial number of products.
	
	\subsection{Low-Rank Tensor MMSE (LR-TMMSE) Filter}
	In this section, we introduce a tensor equalizer which solves the computational complexity issues of the classical MMSE filter. First, assume the number $N$ of antennas at the BS can be factorized as $N = \prod_{d=1}^D N_d$. Then, we reshape the column vector $\bm{w}$ into a $D$-th order tensor $\mc{W} \in \mbb{C}^{N_1 \times \cdots \times N_D}$. The entry $w_n$ of $\bm{w}$ relates to the entry $w_{n_1, \ldots, n_D}$ of $\mc{W}$ as $n= n_1 + (n_2-1)N_1 + \cdots + (n_D-1) \prod_{m=1}^{D-1} N_m$ for $n \in \{1,\ldots, N\}$ and $n_d \in \{1,\ldots,N_d\},\, \forall d \in \{1,\ldots, D\}$. With such a reshaping operation, the equalizer output is rewritten as
	\begin{align}
	y[k] &= \bm{w}^\hermit \bm{x}[k] = \sum_{n=1}^N w_n^* x_n[k] \\
	&=\sum_{n_1,\ldots,n_D=1}^{N_1,\ldots,N_D} w_{n_1,\ldots,n_D}^* x_{n_1,\ldots,n_D}[k], \label{eq:output}
	\end{align}
	where $\mc{X}[k] \in \mbb{C}^{N_1 \times \cdots \times N_D}$ denotes the $D$-dimensional reshape~of~$\bm{x}[k]$.
	
	In previous works~\cite{ribeiro_tensor_2016,ribeiro_iet18}, we consider rank-$1$ separable tensor filters, i.e., $\mc{W}$ is written as an outer product of vectors. However, such a structure is too strict for some applications. For example, the channel matrix \eqref{eq:chan} cannot be separated as a Kronecker product, and, thus, rank-$1$ tensor filters would exhibit poor equalization performance. To overcome this limitation, let us decompose $\mc{W}$ as a sum of $R$ rank-$1$ terms:
	\begin{equation}
	\mc{W} = \sum_{r=1}^R \bm{w}_{1,r} \circ \ldots \circ \bm{w}_{D,r}. \label{eq:tenfilt}
	\end{equation}
	where $\bm{w}_{d,r} \in \mbb{C}^{N_d \times 1}$ for $d \in \{1,\ldots, D\}$, $r \in \{1,\ldots,R\}$ and $R$ denotes the filter rank. Note that \eqref{eq:tenfilt} is known as the canonical polyadic (CP) decomposition in tensor literature~\cite{kolda_tensor_2009,comon2014tensors}. It is interesting to mention the relationship between the outer and Kronecker product notations. Specifically, by vectorizing~\eqref{eq:tenfilt}, we obtain $\text{vec}(\mc{W}) =\sum_{r=1}^R \bm{w}_{D,r} \otimes \ldots \otimes \bm{w}_{1,r}$~\cite{comon2014tensors}. The extra degrees of freedom brought by the $R$ separable components allow the CP tensor to better equalize non-separable systems such as~\eqref{eq:chan}. Note that $R$ is a parameter to be chosen by the filter  designer. As it grows, the filter has more degrees of freedom, but also more parameters to estimate. Therefore a judicious choice of $R$ which balances the performance-complexity trade-off is preferred.
	
	Assuming structure \eqref{eq:tenfilt}, the filter coefficients can  be written as
	\begin{equation}
	w_{n_1,\ldots,n_D} = \sum_{r=1}^R \prod_{d=1}^D [\bm{w}_{d,r}]_{n_d}, \label{eq:elcpd}
	\end{equation}
	which allows us to recast the equalizer output $y[k]$ as follows
	\begin{align}
		y[k] &= \sum_{n_1, \ldots, n_D=1}^{N_1, \ldots, N_D} w_{n_1, \ldots, n_D}^* x_{n_1, \ldots, n_D}[k]\\
		&= \sum_{n_1, \ldots, n_D=1}^{N_1, \ldots, N_D} \left( \sum_{r=1}^R [\bm{w}_{1,r}]_{n_1}^* \ldots [\bm{w}_{D,r}]_{n_D}^* \right) x_{n_1, \ldots, n_D}[k]. \label{eq:oi}
	\end{align}
	
		\begin{algorithm}[t]
		\caption{Low-Rank Tensor MMSE Equalizer}
		\label{alg:lrtmmse}
		\begin{algorithmic}[1]
			\Require{Received signals $\bm{X} \in \mbb{C}^{N \times K}$, training sequence $\bm{s}_u \in \mbb{C}^{K}$, filter rank $R$, filter order $D$, filter dimensions $N_d$ for $d\in\{1,\ldots,D\}$. }
			\State Initialize $\bm{w}_{d,r}$ as $[1,0,\ldots,0]^\tran$, $d \in \{1,\ldots D\}$, $r \in \{1,\ldots, R\}$
			\Repeat
			\For{$d= 1,\ldots,D$}
			\State Build $\bm{U}_{d,r}$ for $r=1,\ldots,R$ by \eqref{eq:moden}
			\State Form $\bm{U}_d = \left[ \bm{U}_{d,1}^\tran,\ldots,\bm{U}_{d,R}^\tran \right]^\tran$
			\State Estimate $\bm{R}_{u_d u_d}$ and $\bm{p}_{u_d}$   by \eqref{eq:modedest}
			\State Update $\bm{w}_d$ by \eqref{eq:tmmsefilt}
			\EndFor 		
			\Until{convergence criterion triggers}
			\State Form tensor filter $\mc{W}$ using \eqref{eq:tenfilt} and \eqref{eq:wd}
			\State $\bm{w} \gets \text{vec}(\mc{W})$
		\end{algorithmic}
	\end{algorithm}
	
	By isolating $\bm{w}_{d,r}$ from the other $(D-1)$ factors, we get:
	\begin{align}
	y[k] & =\sum_{r=1}^{R} \sum_{n_d=1}^{N_d} [\bm{w}_{d,r}]_{n_d}^* \left( \sum_{n_q=1}^{N_q} \prod_{q \neq d}^{D} [\bm{w}_{q,r}]_{n_q}^*x_{n_1, \ldots, n_D}[k]\right)\\
	& =\sum_{r=1}^{R} \sum_{n_d=1}^{N_d} [\bm{w}_{d,r}]_{n_d}^* [\bm{u}_{d,r}[k]]_{n_d} = \bm{w}_{d}^\hermit \bm{u}_d[k] \label{eq:dmodeout}
	\end{align}
	where we define
	\begin{gather}
	\bm{u}_{d,r}[k] = \bm{X}_{(d)}[k]  \bar{\bm{w}}_{d,r}^*\in \mbb{C}^{N_d} \label{eq:dmodeinput1},\\
	\bm{u}_d[k] = \left[ \bm{u}_{d,1}^\tran[k], \ldots, \bm{u}_{d,R}^\tran[k]\right]^\tran \in \mbb{C}^{RN_d} \label{eq:dmodeinput},\\
	\bm{w}_d= \left[ \bm{w}_{d,1}^\tran, \ldots, \bm{w}_{d,R}^\tran \right]^\tran \in \mbb{C}^{RN_d} \label{eq:wd}, \\
	\bar{\bm{w}}_{d,r} = \bigotimes_{q \neq d}^{D} \bm{w}_{q,r} \in \mbb{C}^{\bar{N}_d \times 1},\, \bar{N}_d = \prod_{q \neq d}^D N_q.
	\end{gather}
	Matrix $\bm{X}_{(d)}[k] \in \mbb{C}^{N_d \times \bar{N}_d}$ denotes the $d$-mode matrix unfolding of $\mc{X}[k]$. The tensor element $(n_1,\ldots,n_D)$ maps to the $\bm{X}_{(d)}[k]$ element $(n_d,j)$ as $j=1 + \sum_{j\neq d}^D(n_j-1)\prod_{m\neq d}^{j-1} N_m$~\cite{kolda_tensor_2009}. For more information on tensor notation, the reader is kindly referred to~\cite{kolda_tensor_2009,comon2014tensors}. Equation \eqref{eq:dmodeout} explicits the multilinear property of our tensor filter since $y[k]$ is linear with respect to $\bm{w}_d$ given that $\bar{\bm{w}}_{d,r}$ is fixed for all $d \in \{1,\ldots, D\}$.
	
	The multilinear filter output \eqref{eq:dmodeout} allows us to reformulate the linear equalization problem~\eqref{eq:mse} as
	\begin{equation}
	\min_{\bm{w}_d}\, \esp{ |s_u[k-\delta] - \bm{w}_d^\hermit \bm{u}_d[k]|^2 },\quad d \in \{1,\ldots, D\}.  \label{eq:tmmseprob}
	\end{equation}
	There are several ways to solve~\eqref{eq:tmmseprob}. We propose an alternating minimization approach in which we solve for each $\bm{w}_d$ sequentially until a convergence criterion is satisfied. For a given $d$, \eqref{eq:tmmseprob} can be seen as a low-dimensional MMSE problem. Hence, we have that
	\begin{gather}
	\bm{w}_{d,\text{MMSE}} = \bm{R}_{u_d, u_d}^{-1} \bm{p}_{u_d} \in \mbb{C}^{RN_d} \label{eq:tmmsefilt},\\
	\bm{R}_{u_d, u_d} = \esp{\bm{u}_{d}[k]\bm{u}_d^\hermit[k]}\in \mbb{C}^{RN_d \times RN_d} \label{eq:covmtx},\\
	\bm{p}_{u_d} = \esp{ \bm{u}_d[k] s_u^*[k-\delta] }  \in \mbb{C}^{RN_d}  \label{eq:crosscovvec}
	\end{gather}
	for $d \in \{1,\ldots, D\}$.
	
	Let us calculate the sample estimates of $\bm{R}_{u_d u_d}$ and $\bm{p}_{u_d}$. To this end, define $\mc{X} \in \mbb{C}^{N_1 \times \cdots \times N_D \times K}$ as the tensor reshaping of~$\bm{X}$. Now, consider the following tensor-vector product~\cite{kolda_tensor_2009} evaluated at the tensor modes $\mbb{J}_d = \{ j = 1,\ldots,D \,\vert\, j \neq d \}$
	\begin{equation}
	\bm{U}_{d,r} = \mc{X} \bigtimes_{j \in \mbb{J}_d} \bm{w}_{j,r}^\hermit \in \mbb{C}^{N_d \times K}. \label{eq:moden}
	\end{equation}
	The elements of \eqref{eq:moden} are given by
	\begin{align}
	&[\bm{U}_{d,r}]_{n_d,k} = \\
	&\sum_{n_1=1}^{N_1} \ldots \sum_{n_{d-1}=1}^{N_{d-1}} \sum_{n_{d+1}=1}^{N_{d+1}} \ldots \sum_{n_{D}=1}^{N_D} [\mc{X}]_{n_1,\ldots,n_D,k} \prod_{j\in \mbb{J}_d}
	[\bm{w}_{d,r}]_{n_j}^*.
	\end{align}
	Define $\bm{U}_d = \left[ \bm{U}_{d,1}^\tran,\ldots,\bm{U}_{d,R}^\tran \right]^\tran \in \mbb{C}^{RN_d \times K}$. The statistics \eqref{eq:covmtx} and \eqref{eq:crosscovvec} may be estimated as
	\begin{equation}
	\bm{R}_{u_d u_d} \approx \frac{1}{K} \bm{U}_d \bm{U}_d^\hermit,\quad\bm{p}_{u_d} \approx \frac{1}{K}\bm{U}_d \bm{s}_u^* \label{eq:modedest}
	\end{equation}
	for $d \in \{1,\ldots,D\}$. Henceforth, this method is referred to as low-rank tensor MMSE (LR-TMMSE) filter and it is summarized in Algorithm~\ref{alg:lrtmmse}.
	
	The LR-TMMSE equalizer is an iterative method. Let us assume that it converges within $I$ iterations. Each iteration carries out $R(D-1)NK$ products to compute $\bm{U}_{d}$,  $N_d^2K+ N_d K$ to estimate the statistics and $O(N_d^3) + N_d^2$ for each $\bm{w}_d$. Therefore, LR-TMMSE carries out a total of
	\begin{align}
	&P_\text{LR-TMMSE}(\{N_d\}, D, I, K) = \label{eq:count_tmmse}\\
	& I \left[ \sum_{d=1}^D R(D-1)N K + N_d^2K + N_dK + O(N_d^3) + N_d \right]
	\end{align}
	products.

	\section{Simulation Results}
		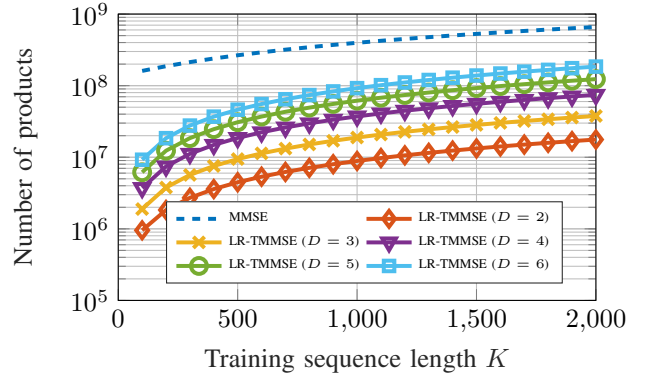
\begin{figure}[t]
		\centering
\definecolor{mycolor1}{rgb}{0.00000,0.44700,0.74100}%
\definecolor{mycolor2}{rgb}{0.85000,0.32500,0.09800}%
\definecolor{mycolor3}{rgb}{0.92900,0.69400,0.12500}%
\definecolor{mycolor4}{rgb}{0.49400,0.18400,0.55600}%
\definecolor{mycolor5}{rgb}{0.46600,0.67400,0.18800}%
\definecolor{mycolor6}{rgb}{0.30100,0.74500,0.93300}%
\begin{tikzpicture}

\begin{axis}[%
width=2.5in,
height=1.5in,
scale only axis,
xmin=0,
xmax=2000,
xlabel style={font=\color{white!15!black}},
xlabel={Training sequence length $K$},
ymode=log,
ymin=100000,
ymax=1000000000,
yminorticks=true,
ylabel style={font=\color{white!15!black}},
ylabel={Number of products},
axis background/.style={fill=white},
xmajorgrids,
ymajorgrids,
yminorgrids,
legend columns = 2,
legend style={font=\tiny, at={(0.1,0.059)}, anchor=south west, legend cell align=left, align=left, draw=white!15!black}
]
\addplot [color=mycolor1, dashed, line width=1.5pt,mark size=3pt]
  table[row sep=crcr]{%
100	160745472\\
200	187011072\\
300	213276672\\
400	239542272\\
500	265807872\\
600	292073472\\
700	318339072\\
800	344604672\\
900	370870272\\
1000	397135872\\
1100	423401472\\
1200	449667072\\
1300	475932672\\
1400	502198272\\
1500	528463872\\
1600	554729472\\
1700	580995072\\
1800	607260672\\
1900	633526272\\
2000	659791872\\
};
\addlegendentry{MMSE}

\addplot [color=mycolor2, line width=1.5pt, mark=diamond, mark options={solid, mycolor2},mark size=3pt]
  table[row sep=crcr]{%
100	953824\\
200	1833824\\
300	2713824\\
400	3593824\\
500	4473824\\
600	5353824\\
700	6233824\\
800	7113824\\
900	7993824\\
1000	8873824\\
1100	9753824\\
1200	10633824\\
1300	11513824\\
1400	12393824\\
1500	13273824\\
1600	14153824\\
1700	15033824\\
1800	15913824\\
1900	16793824\\
2000	17673824\\
};
\addlegendentry{LR-TMMSE ($D=2$)}

\addplot [color=mycolor3, line width=1.5pt, mark=x, ,mark size=3pt, mark options={solid, mycolor3}]
  table[row sep=crcr]{%
100	1889520\\
200	3775920\\
300	5662320\\
400	7548720\\
500	9435120\\
600	11321520\\
700	13207920\\
800	15094320\\
900	16980720\\
1000	18867120\\
1100	20753520\\
1200	22639920\\
1300	24526320\\
1400	26412720\\
1500	28299120\\
1600	30185520\\
1700	32071920\\
1800	33958320\\
1900	35844720\\
2000	37731120\\
};
\addlegendentry{LR-TMMSE ($D=3$)}

\addplot [color=mycolor4, line width=1.5pt, mark=triangle, ,mark size=3pt, mark options={solid, rotate=180, mycolor4}]
  table[row sep=crcr]{%
100	3722636\\
200	7443036\\
300	11163436\\
400	14883836\\
500	18604236\\
600	22324636\\
700	26045036\\
800	29765436\\
900	33485836\\
1000	37206236\\
1100	40926636\\
1200	44647036\\
1300	48367436\\
1400	52087836\\
1500	55808236\\
1600	59528636\\
1700	63249036\\
1800	66969436\\
1900	70689836\\
2000	74410236\\
};
\addlegendentry{LR-TMMSE ($D=4$)}

\addplot [color=mycolor5, line width=1.5pt, mark=o, ,mark size=3pt, mark options={solid, mycolor5}]
  table[row sep=crcr]{%
100	6170152\\
200	12338952\\
300	18507752\\
400	24676552\\
500	30845352\\
600	37014152\\
700	43182952\\
800	49351752\\
900	55520552\\
1000	61689352\\
1100	67858152\\
1200	74026952\\
1300	80195752\\
1400	86364552\\
1500	92533352\\
1600	98702152\\
1700	104870952\\
1800	111039752\\
1900	117208552\\
2000	123377352\\
};
\addlegendentry{LR-TMMSE ($D=5$)}

\addplot [color=mycolor6, line width=1.5pt, mark size=2pt, mark=square, mark options={solid, mycolor6}]
  table[row sep=crcr]{%
100	9232068\\
200	18463668\\
300	27695268\\
400	36926868\\
500	46158468\\
600	55390068\\
700	64621668\\
800	73853268\\
900	83084868\\
1000	92316468\\
1100	101548068\\
1200	110779668\\
1300	120011268\\
1400	129242868\\
1500	138474468\\
1600	147706068\\
1700	156937668\\
1800	166169268\\
1900	175400868\\
2000	184632468\\
};
\addlegendentry{LR-TMMSE ($D=6$)}

\end{axis}
\end{tikzpicture}%
		\caption{$N=512$ antennas, $I=2$ iterations, filter rank $R=3$.}
		\label{fig:comp-samples}
	\end{figure}
	~
	\begin{figure}[t]
		\centering
\definecolor{mycolor1}{rgb}{0.00000,0.44700,0.74100}%
\definecolor{mycolor2}{rgb}{0.85000,0.32500,0.09800}%
\definecolor{mycolor3}{rgb}{0.92900,0.69400,0.12500}%
\definecolor{mycolor4}{rgb}{0.49400,0.18400,0.55600}%
\begin{tikzpicture}

\begin{axis}[%
width=2.5in,
height=1.5in,
scale only axis,
xmin=0,
xmax=2048,
xlabel style={font=\color{white!15!black}},
xlabel={Number $N$ of antennas},
ymode=log,
ymin=1000,
ymax=100000000000,
yminorticks=true,
ylabel style={font=\color{white!15!black}},
ylabel={Number of products},
axis background/.style={fill=white},
xmajorgrids,
ymajorgrids,
yminorgrids,
legend columns = 2,
legend style={font=\tiny,at={(0.10,0.085)}, anchor=south west, legend cell align=left, align=left, draw=white!15!black}
]
\addplot [color=mycolor1, dashed, line width=1.5pt]
  table[row sep=crcr]{%
16	167552\\
32	667392\\
64	2762240\\
128	12020736\\
256	56317952\\
512	292073472\\
1024	1704550400\\
2048	11111940096\\
};
\addlegendentry{MMSE}

\addplot [color=mycolor2, line width=1.5pt, mark=diamond, mark size=3pt, mark options={solid, mycolor2}]
  table[row sep=crcr]{%
16	4272\\
32	15524\\
64	57240\\
128	226592\\
256	872616\\
512	3370032\\
1024	13445184\\
2048	52872272\\
};
\addlegendentry{LR-TMMSE ($R=1$)}

\addplot [color=mycolor3, line width=1.5pt, mark=x, mark size=3pt, mark options={solid, mycolor3}]
  table[row sep=crcr]{%
16	7344\\
32	27812\\
64	106392\\
128	423200\\
256	1659048\\
512	6515760\\
1024	26028096\\
2048	103203920\\
};
\addlegendentry{LR-TMMSE ($R=2$)}

\addplot [color=mycolor4, line width=1.5pt, mark=triangle, mark size=3pt, mark options={solid, rotate=180, mycolor4}]
  table[row sep=crcr]{%
16	10416\\
32	40100\\
64	155544\\
128	619808\\
256	2445480\\
512	9661488\\
1024	38611008\\
2048	153535568\\
};
\addlegendentry{LR-TMMSE ($R=3$)}

\end{axis}
\end{tikzpicture}%
		\caption{$K=600$ symbols, $I=2$ iterations, filter order $D=3$.}
		\label{fig:comp-array}
	\end{figure}
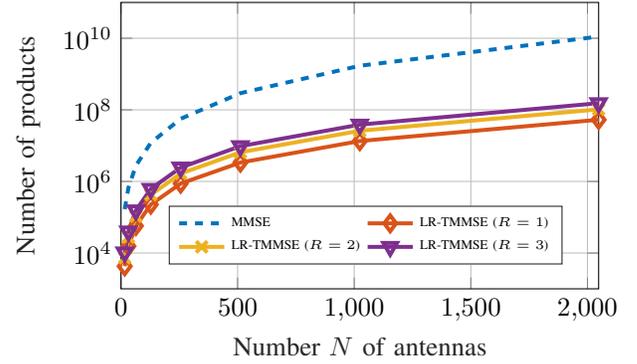
	
	In this section, we present simulation results conducted to analyze the performance of the proposed LR-TMMSE equalizer. In all simulations, the sample-based MMSE is considered as the benchmark equalizer. We consider two figures of merit: (i) number of products to obtain the filter weights vector according to \eqref{eq:count_mmse} and \eqref{eq:count_tmmse}, (ii) the signal to noise and interference ratio (SINR) after  equalization. For a given equalizer $\bm{w}$, we have $\text{SINR}(\bm{w}) = (\bm{w}^\hermit \bm{R}_{xx} \bm{w})/(\bm{w}^\hermit (\bm{R}_{ii} + \bm{R}_{bb}) \bm{w})$. The figures presented in this section were obtained by averaging the results from $1000$ independent experiments. Each experiment consists of generating a $K$-length QPSK-modulated symbol sequence for all $U$ UE, building a channel realization, forming the observed signals and finally applying the equalizers. The desired signal delay $\delta$ is optimized for all equalizers as explained in Sec.~\ref{sec:mmse}. We consider the following parameter setup in our simulations: $U=4$ users, $\sigma_s^2=1$, $L=5$ channel paths, the directions of arrival $\theta_{u,\ell}$ are drawn from a random variable uniformly distributed in $[-90^\circ, 90^\circ]$, the sinc function is set as the effective pulse-shaping waveform $g(t)$. LR-TMMSE achieves convergence when $\| \bm{w}_{i+1} - \bm{w}_i\|_2^2 < \epsilon $, where $i$ denotes the iteration number and $\epsilon$ is a small positive threshold. We set $\epsilon=0.1$ and, according to a preliminary simulation, the algorithm typically converges within $I=2$ iterations.
	
	The computational complexity of LR-TMMSE is studied in Figures~\ref{fig:comp-samples} and~\ref{fig:comp-array}. In the former figure, we investigate the influence of the training sequence length $K$ on the number of products for tensor equalizers with different order $D$. We fix the number of antennas to $N=512$ and we consider $D$-order tensor filters such that their dimensions $\{N_d\}$ satisfy $\prod_{d=1}^{D} N_d = 512$. Figure~\ref{fig:comp-samples} reveals that LR-TMMSE computes fewer products than the benchmark for the considered parameters. Moreover, it shows that the complexity of LR-TMMSE increases with $D$. Although the cubic term in \eqref{eq:count_tmmse} tends to become less important as $D$ grows, we observe a significant overhead associated with the computation of the $\bm{U}_d$ matrices. We plot the computational complexity as a function of the BS array size $N$ for different ranks $R$ in Figure~\ref{fig:comp-array}. We observe that $R$ does not influence much on the complexity. Besides, LR-TMMSE is more computationally efficient than the benchmark even for very large array sizes.
	
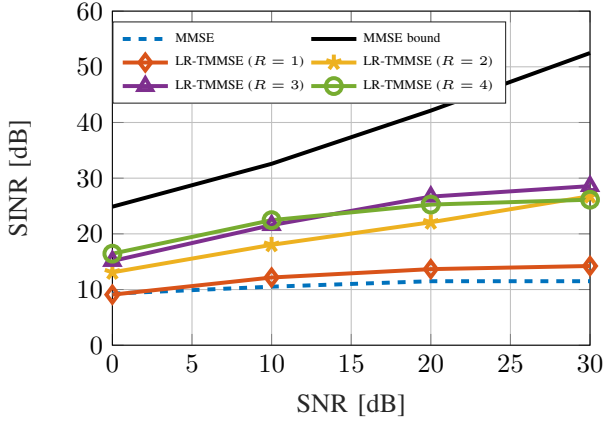
\begin{figure}[t]
		\centering
\definecolor{mycolor1}{rgb}{0.00000,0.44700,0.74100}%
\definecolor{mycolor2}{rgb}{0.85000,0.32500,0.09800}%
\definecolor{mycolor3}{rgb}{0.92900,0.69400,0.12500}%
\definecolor{mycolor4}{rgb}{0.49400,0.18400,0.55600}%
\definecolor{mycolor5}{rgb}{0.46600,0.67400,0.18800}%
\begin{tikzpicture}

\begin{axis}[%
width=2.5in,
height=1.75in,
scale only axis,
xmin=0,
xmax=30,
xlabel style={font=\color{white!15!black}},
xlabel={SNR [dB]},
ymin=0,
ymax=60,
ylabel style={font=\color{white!15!black}},
ylabel={SINR [dB]},
axis background/.style={fill=white},
xmajorgrids,
ymajorgrids,
legend columns = 2,
ytick distance = 10,
legend style={font=\tiny, at={(0.0,0.723)}, anchor=south west, legend cell align=left, align=left, draw=white!15!black}
]
\addplot [color=mycolor1, dashed, mark options={solid, mycolor1}, line width = 1.5pt]
  table[row sep=crcr]{%
0	9.26927991865948\\
10	10.5124585781969\\
20	11.4942541388903\\
30	11.4954635724321\\
};
\addlegendentry{MMSE}

\addplot [color=black, line width = 1.5pt]
  table[row sep=crcr]{%
0	24.8791957594416\\
10	32.5812679570945\\
20	42.1291830220341\\
30	52.4762246993276\\
};
\addlegendentry{MMSE bound}

\addplot [color=mycolor2, mark=diamond, mark size=3pt, mark options={solid, mycolor2}, line width = 1.5pt]
  table[row sep=crcr]{%
0	9.0801887639269\\
10	12.1602916727273\\
20	13.6647236947568\\
30	14.2260926124159\\
};
\addlegendentry{LR-TMMSE ($R=1$)}

\addplot [color=mycolor3, mark=star, mark size=3pt, mark options={solid, mycolor3}, line width = 1.5pt]
  table[row sep=crcr]{%
0	13.0841572514254\\
10	18.028330834712\\
20	22.0789680665001\\
30	26.766485771479\\
};
\addlegendentry{LR-TMMSE ($R=2$)}

\addplot [color=mycolor4, mark=triangle, mark size=3pt, mark options={solid, mycolor4}, line width = 1.5pt]
  table[row sep=crcr]{%
0	15.1244953062607\\
10	21.5691970116857\\
20	26.686610302766\\
30	28.5766879337789\\
};
\addlegendentry{LR-TMMSE ($R=3$)}

\addplot [color=mycolor5, mark=o, mark size=3pt, mark options={solid, mycolor5}, line width = 1.5pt]
  table[row sep=crcr]{%
0	16.4370153005811\\
10	22.4586321747981\\
20	25.2547402318792\\
30	26.1337052927009\\
};
\addlegendentry{LR-TMMSE ($R=4$)}

\end{axis}
\end{tikzpicture}%
		\caption{$N=512$ antennas, $K=600$ symbols, filter order $D=3$.}
		\label{fig:snr_R}
	\end{figure}

\begin{figure}[t]
		\centering
\definecolor{mycolor1}{rgb}{0.00000,0.44700,0.74100}%
\definecolor{mycolor2}{rgb}{0.85000,0.32500,0.09800}%
\definecolor{mycolor3}{rgb}{0.92900,0.69400,0.12500}%
\definecolor{mycolor4}{rgb}{0.49400,0.18400,0.55600}%
\definecolor{mycolor5}{rgb}{0.46600,0.67400,0.18800}%
\begin{tikzpicture}

\begin{axis}[%
width=2.5in,
height=1.75in,
scale only axis,
xmin=0,
xmax=30,
xlabel style={font=\color{white!15!black}},
xlabel={SNR [dB]},
ymin=0,
ymax=60,
ylabel style={font=\color{white!15!black}},
ylabel={SINR [dB]},
axis background/.style={fill=white},
xmajorgrids,
ymajorgrids,
legend columns=2,
ytick distance=10,
legend style={font=\tiny, at={(0,0.727)}, anchor=south west, legend cell align=left, align=left, draw=white!15!black}
]
\addplot [color=mycolor1, dashed, mark options={solid, mycolor1}, line width = 1.5pt]
  table[row sep=crcr]{%
0	9.26927991865948\\
10	10.5124585781969\\
20	11.4942541388903\\
30	11.4954635724321\\
};
\addlegendentry{MMSE}

\addplot [color=black, line width = 1.5pt]
  table[row sep=crcr]{%
0	24.8791957594416\\
10	32.5812679570945\\
20	42.1291830220341\\
30	52.4762246993276\\
};
\addlegendentry{MMSE bound}

\addplot [color=mycolor2, mark options={solid, mycolor2}, mark size=3pt, mark=diamond, line width=1.5pt]
  table[row sep=crcr]{%
0	17.0446187665711\\
10	21.4283731596294\\
20	22.7938461370955\\
30	23.0066728866679\\
};
\addlegendentry{LR-TMMSE ($D=2$)}

\addplot [color=mycolor3, mark=asterisk, mark size=3pt, mark options={solid, mycolor3}, line width=1.5pt]
  table[row sep=crcr]{%
0	14.9986494594761\\
10	21.523863532466\\
20	26.4899485147611\\
30	28.6891443572073\\
};
\addlegendentry{LR-TMMSE ($D=3$)}

\addplot [color=mycolor4, mark=triangle, mark size=3pt, mark options={solid, mycolor4}, line width=1.5pt]
  table[row sep=crcr]{%
0	14.2879412513937\\
10	20.2247190964576\\
20	25.976714747273\\
30	29.48812665339\\
};
\addlegendentry{LR-TMMSE ($D=4$)}

\addplot [color=mycolor5, mark=o, mark size=3pt, mark options={solid, mycolor5}, line width=1.5pt]
  table[row sep=crcr]{%
0	12.9001166652436\\
10	17.4537564650608\\
20	22.4264898063972\\
30	28.708058153422\\
};
\addlegendentry{LR-TMMSE ($D=5$)}

\end{axis}
\end{tikzpicture}%
		\caption{$N=512$ antennas, $K=600$ symbols, filter rank $R=3$.}
		\label{fig:snr_D}
	\end{figure}
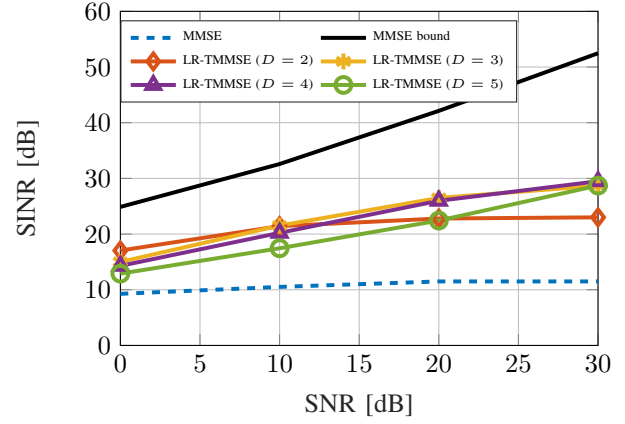

	We investigate the SINR performance for different ranks $R$ and orders $D$ for only $K=600$ training symbols in Figures~\ref{fig:snr_R} and~\ref{fig:snr_D}. We consider the theoretical MMSE as an SINR upper bound. Figure~\ref{fig:snr_R} shows that $R=1$ performs poorly, as expected. This is because the signal model \eqref{eq:rx} is not separable. Therefore, it is necessary to increase the filter rank $R$ to better equalize the system. However, we note that SINR is rather similar for $R=3$ and $R=4$, indicating that the LR-TMMSE performance is bounded with respect to this parameter. We plot the SINR for $R=3$ and different $D$ in Figure~\ref{fig:snr_D}. This result suggests that the tensor order has a limited effect on the SINR. Still, the difference between $D=2$ and $D \in \{3,4,5\}$ at $30$ dB SNR is $5$ dB.
	
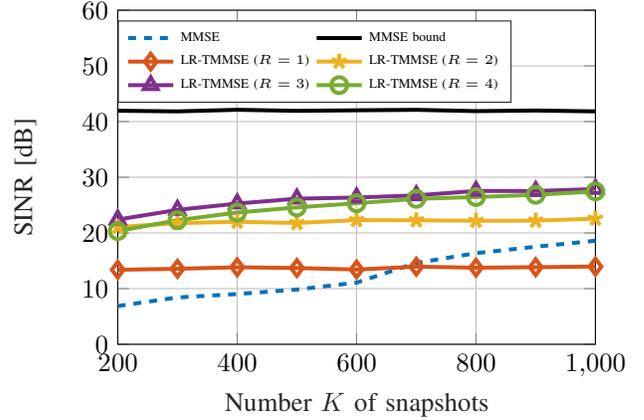
\begin{figure}[t]
		\centering
\definecolor{mycolor1}{rgb}{0.00000,0.44700,0.74100}%
\definecolor{mycolor2}{rgb}{0.85000,0.32500,0.09800}%
\definecolor{mycolor3}{rgb}{0.92900,0.69400,0.12500}%
\definecolor{mycolor4}{rgb}{0.49400,0.18400,0.55600}%
\definecolor{mycolor5}{rgb}{0.46600,0.67400,0.18800}%
\begin{tikzpicture}

\begin{axis}[%
width=2.5in,
height=1.75in,
scale only axis,
xmin=200,
xmax=1000,
xlabel style={font=\color{white!15!black}},
xlabel={Number $K$ of snapshots},
ymin=0,
ymax=60,
ylabel style={font=\color{white!15!black}},
ylabel={SINR [dB]},
axis background/.style={fill=white},
xmajorgrids,
ymajorgrids,
ytick distance=10,
legend columns = 2,
legend style={font=\tiny, at={(0,0.723)}, anchor=south west, legend cell align=left, align=left, draw=white!15!black}
]
\addplot [color=mycolor1, dashed, mark options={solid, mycolor1}, mark size=3pt, line width = 1.5pt]
  table[row sep=crcr]{%
200	6.87236559466678\\
300	8.4193317274454\\
400	9.02440339334513\\
500	9.83426429807215\\
600	11.0793549230624\\
700	14.6090096247955\\
800	16.368351983351\\
900	17.5251932827834\\
1000	18.5985245903872\\
};
\addlegendentry{MMSE}

\addplot [color=black, line width = 1.5pt]
  table[row sep=crcr]{%
200	41.9491102013604\\
300	41.8308746661446\\
400	42.1187798039687\\
500	41.9236696842706\\
600	42.0268220652022\\
700	42.1083238175893\\
800	41.8782748011208\\
900	41.9839631965811\\
1000	41.8332766986375\\
};
\addlegendentry{MMSE bound}

\addplot [color=mycolor2, line width = 1.5pt, mark size=3pt, mark=diamond]
  table[row sep=crcr]{%
200	13.3828391103292\\
300	13.5625522973867\\
400	13.8350482529334\\
500	13.7009062721414\\
600	13.4263261916369\\
700	13.942227738665\\
800	13.7570526507847\\
900	13.8480484741177\\
1000	13.9645003272541\\
};
\addlegendentry{LR-TMMSE ($R=1$)}

\addplot [color=mycolor3, line width = 1.5pt, mark size=3pt, mark=star]
  table[row sep=crcr]{%
200	21.0589743428419\\
300	21.760637717283\\
400	22.0114237793931\\
500	21.7781884123828\\
600	22.3151465130413\\
700	22.2683834909589\\
800	22.1817945670897\\
900	22.2076627435633\\
1000	22.572425523071\\
};
\addlegendentry{LR-TMMSE ($R=2$)}

\addplot [color=mycolor4, line width = 1.5pt, mark size=3pt, mark=triangle]
  table[row sep=crcr]{%
200	22.3960856907801\\
300	24.1412217992422\\
400	25.2778318139333\\
500	26.1527443463435\\
600	26.3572439125574\\
700	26.7450152209928\\
800	27.5392711246146\\
900	27.5205931354318\\
1000	27.8879353132474\\
};
\addlegendentry{LR-TMMSE ($R=3$)}

\addplot [color=mycolor5, line width = 1.5pt, mark size=3pt, mark=o]
  table[row sep=crcr]{%
200	20.3278042784792\\
300	22.2887121636307\\
400	23.6582384371518\\
500	24.5874797167132\\
600	25.3337690507839\\
700	26.1199105883607\\
800	26.4453693928498\\
900	26.8416663905896\\
1000	27.4535424551274\\
};
\addlegendentry{LR-TMMSE ($R=4$)}

\end{axis}
\end{tikzpicture}%
		\caption{$N=512$ antennas, SNR = $20$ dB, filter order $D=3$.}
		\label{fig:sampR}
	\end{figure}

\begin{figure}[t]
		\centering
\definecolor{mycolor1}{rgb}{0.00000,0.44700,0.74100}%
\definecolor{mycolor2}{rgb}{0.85000,0.32500,0.09800}%
\definecolor{mycolor3}{rgb}{0.92900,0.69400,0.12500}%
\definecolor{mycolor4}{rgb}{0.49400,0.18400,0.55600}%
\definecolor{mycolor5}{rgb}{0.46600,0.67400,0.18800}%
\begin{tikzpicture}

\begin{axis}[%
width=2.5in,
height=1.75in,
scale only axis,
xmin=200,
xmax=1000,
xlabel style={font=\color{white!15!black}},
xlabel={Number $K$ of snapshots},
ymin=0,
ymax=60,
ylabel style={font=\color{white!15!black}},
ylabel={SINR [dB]},
axis background/.style={fill=white},
xmajorgrids,
ymajorgrids,
ytick distance=10,
legend columns=2,
legend style={font=\tiny,at={(0,0.723)}, anchor=south west, legend cell align=left, align=left, draw=white!15!black}
]
\addplot [color=mycolor1, dashed, mark size=3pt, mark options={solid, mycolor1}, line width=1.5pt]
  table[row sep=crcr]{%
200	6.87236559466678\\
300	8.4193317274454\\
400	9.02440339334513\\
500	9.83426429807215\\
600	11.0793549230624\\
700	14.6090096247955\\
800	16.368351983351\\
900	17.5251932827834\\
1000	18.5985245903872\\
};
\addlegendentry{MMSE}

\addplot [color=black, line width=1.5pt]
  table[row sep=crcr]{%
200	41.9491102013604\\
300	41.8308746661446\\
400	42.1187798039687\\
500	41.9236696842706\\
600	42.0268220652022\\
700	42.1083238175893\\
800	41.8782748011208\\
900	41.9839631965811\\
1000	41.8332766986375\\
};
\addlegendentry{MMSE bound}

\addplot [color=mycolor2, line width=1.5pt, mark size=3pt, mark=diamond]
  table[row sep=crcr]{%
200	15.3138392416312\\
300	18.5425572527113\\
400	20.4367549937247\\
500	21.7914936027504\\
600	22.5831941450084\\
700	23.4612954913494\\
800	24.0389883335475\\
900	24.7097000129438\\
1000	25.2660786528366\\
};
\addlegendentry{LR-TMMSE ($D=2$)}

\addplot [color=mycolor3, line width=1.5pt, mark size=3pt, mark=star]
  table[row sep=crcr]{%
200	22.534067121091\\
300	24.3500503639626\\
400	25.1899957257652\\
500	26.0260852138968\\
600	26.5663513045453\\
700	27.1441923861867\\
800	27.2235036630242\\
900	27.4774619047479\\
1000	27.7262308375025\\
};
\addlegendentry{LR-TMMSE ($D=3$)}

\addplot [color=mycolor4, line width=1.5pt, mark size=3pt, mark=triangle]
  table[row sep=crcr]{%
200	22.9523573728555\\
300	24.1438048532508\\
400	25.1644142597554\\
500	25.7172483405223\\
600	26.0176714562021\\
700	26.1968023852977\\
800	26.4368395970953\\
900	26.7365315091115\\
1000	26.9000305177616\\
};
\addlegendentry{LR-TMMSE ($D=4$)}

\addplot [color=mycolor5, line width=1.5pt, mark size=3pt, mark=o]
  table[row sep=crcr]{%
200	21.3048774713772\\
300	21.7625659038154\\
400	22.2075851669241\\
500	22.2562791311864\\
600	22.5056750423418\\
700	22.5245578357285\\
800	22.4072451949843\\
900	22.5712287691816\\
1000	22.6150268490368\\
};
\addlegendentry{LR-TMMSE ($D=5$)}

\end{axis}
\end{tikzpicture}%
		\caption{$N=512$ antennas,  SNR = $20$ dB, filter rank $R=3$.}
		\label{fig:sampD}
	\end{figure}
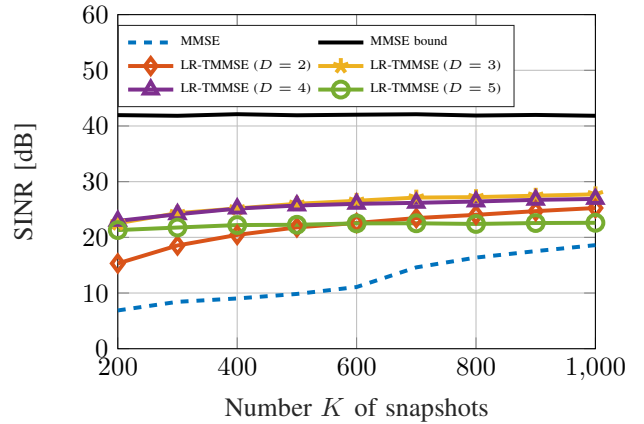

	We finally analyze the effects of the training sequence length $K$ on the SINR for $N=512$ antennas in Figures~\ref{fig:sampR} and \ref{fig:sampD}. LR-TMMSE yields its worst performance when $R=1$. In this case, even the benchmark performs better for $K\geq 700$. However, when we increase the filter rank, we notice an SINR gain, which becomes bounded at $R=4$, confirming the discussion in the previous paragraph. In Figure~\ref{fig:sampD}, we notice that SINR slightly varies for different $D$. Yet, LR-TMMSE provides the worst SINR for short sequences with $D=2$.

	~
	~
	~

	\section{Conclusion}
	In this paper, we introduced a novel low-rank tensor MMSE equalizer based on the MMSE filter for non-separable (low-rank) MIMO channels. Our simulation results indicate that the proposed method is more robust to short training symbol sequences and more computationally efficient than the benchmark (linear MMSE) solution. The obtained results also show that the tensor equalizer performance increases with the filter rank up to a certain level. The tensor filter order has shown to be less relevant to the equalizer SINR, although third-order filters yield the best performance.
	
	\cleardoublepage
	\bibliographystyle{IEEEtran}
	\bibliography{./iswcs19}

\end{document}